\documentclass[12pt,english,superscriptaddress,aps,prd,preprint,showkeys,nofootinbib]{revtex4-1}
\usepackage{amsmath}
\usepackage{amssymb}
\usepackage{amsbsy}
\usepackage{amsfonts}
\usepackage{amsopn}
\usepackage{amstext}
\usepackage{graphicx}
\usepackage{amssymb}
\usepackage{amsfonts}
\usepackage{amsmath}
\usepackage{graphicx}
\usepackage[english]{babel}
\usepackage{color}
\usepackage{slashed}
\usepackage{esint}
\usepackage[dvips]{epsfig}
\usepackage[dvips]{graphicx}
\usepackage{float}
\usepackage{units}
\usepackage{textcomp}
\newcommand{\beq}{\begin{equation}}
\newcommand{\eeq}{\end{equation}}
\newcommand{\bea}{\begin{eqnarray}}
\newcommand{\eea}{\end{eqnarray}}

\begin{document}

\title{Ellis-Bronnikov Wormholes in Asymptotically Safe Gravity}

\author{G. Alencar \footnote{geova@fisica.ufc.br}}

\affiliation{${}^{*}$\!Universidade Federal do Cear\'a, Fortaleza-CE, Brazil.}
\affiliation{${}^{}$\!International Institute of Physics - Federal University of Rio Grande do Norte, Campus Universit\'ario, Lagoa Nova, Natal, RN 59078-970, Brazil.}

\author{V. B. Bezerra\footnote{E-mail:valdir@fisica.ufpb.br}}

\affiliation{${}^{\dag}$\!Universidade Federal da Para\'iba, Departamento de F\'isica, Jo\~ao Pessoa, PB,Brazil.}

\author{C. R. Muniz\footnote{E-mail:celio.muniz@uece.br}}

\affiliation{${}^{\ddag}$\!Universidade Estadual do Cear\'a, Faculdade de Educa\c c\~ao, Ci\^encias e Letras de Iguatu, Iguatu-CE, 63.500-000, Brazil.}

\author{H. S. Vieira\footnote{E-mail:horacio.santana.vieira@hotmail.com}}

\affiliation{${}^{\S}$\!Theoretical Astrophysics, Institute for Astronomy and Astrophysics,
University of T\"{u}bingen, 72076 T\"{u}bingen, Germany}

\begin{abstract}

\vspace{0.75cm}
In this paper, we investigate the simplest wormhole solution - the Ellis-Bronnikov one - in the context of the Asymptotically Safe Gravity (ASG) at the Planck scale.  We work with three models, which employ Ricci scalar, Kretschmann scalar, and squared Ricci tensor to improve the field equations by turning the Newton constant into a running coupling constant. For all the cases, we check the radial energy conditions of the wormhole solution and compare them with those valid in General Relativity (GR). We verify that asymptotic safety guarantees that the Ellis-Bronnikov wormhole can satisfy the radial energy conditions at the throat radius, $r_0$, within an interval of values of this latter. That is quite different from the result found in GR. Following, we evaluate the effective radial state parameter, $\omega(r)$, at $r_0$, showing that the quantum gravitational effects modify Einstein's field equations in such a way that it is necessary a very exotic source of matter to generate the wormhole spacetime -- phantom or quintessence-like. That occurs within some ranges of throat radii, even though the energy conditions are or not violated there. Finally, we find that, although at $r_0$ we have a quintessence-like matter, on growing of $r$ we necessarily come across phantom-like regions. We speculate if such a phantom fluid must always be present in wormholes in the ASG context or even in more general quantum gravity scenarios.

\end{abstract}

\renewcommand{\thesection}      {\Roman{section}}
\maketitle
\newpage

\section{Introduction}
Wormholes and their traversability are an object of intense discussion in the communities which study General Relativity and its extensions. By means of Einstein's theory, one shows \cite{Ellis, Bronnikov, Morris, Cataldo} that the wormhole is traversable only with the presence of exotic matter, including Casimir energy \cite{Gar, Jusufi, Alencar:2021ejd,Oliveira:2021ypz,Martinez:2020hjm}, being even capable to mimic the behavior of black holes \cite{Nandi, Karimov, Rosa}. Classically modified gravity theories involving wormholes (e.g. Einstein--Gauss--Bonett gravity \cite{EGB-G}, Lovelock gravity \cite{L-G}, Einstein--Born--Infeld gravity \cite{EBI-G}, and others \cite{Oth-ext,Chew:2016epf,Chew:2018vjp}), as well as some quantum corrected ones\cite{Q-C}, are also quite discussed in the literature.

Nobody knows if quantum effects can change the energy conditions of wormholes and avoid the necessity of having non-exotic matter, but a definitive answer to these issues requires the formulation of the ultimate quantum theory of gravity, which is actually intensely searched. Notwithstanding, quantum effects in gravity can be described through an asymptotically safe quantum field theory, which is UV complete \cite{as}. The existence of such a fixed point for the gravity renormalization group flow is verified from several methods and in various scenarios \cite{Reuter-1st,Lausher,Litim,Machado-11,Benedetti-13,Manrique-20,Christiansen-24,Morris-32,Demmel-36,Platania-40,Cristiansen-43,Falls-45,Narain-49,Oda-54,Eichorn-61,Eichorn-70,Reichert-71,Daas-73}.
Its physical applications are explored in \cite{Bonanno-86,Bonanno-88,Bonanno-89,Bonanno-90,Platania-92,Platania-93,Bonanno-98,Adeifeoba-99,Platania-100}. However, solving the exact renormalization group equation to derive the effective average action is very hard, if not impossible. Therefore the effects of this quantization method are usually considered, ({\it i.e.}, semiclassically) as a correction to the classical theory and studied by means of an effective theory obtained by turning the classical coupling constant into a running one, which is derived from the solution for the $\beta$-function \cite{Reuter-1st, Reuter-2nd, Reuter-3rd}.

An example of the renormalization group improvement of the field equations can be made via the action modification presented in \cite{Reuter-00,our3}. In this method the action functional is covariantly improved leading to the modification of Einstein's equations
\begin{equation}
G_{\mu\nu} = 8\pi G(\chi) T_{\mu\nu} + G(\chi) X_{\mu\nu} (\chi) \ ,\label{IEM}
\end{equation}
where $G(\chi)$ is the improved coupling constant introduced as a function of the curvature invariants $\chi$, with the covariant tensor $X_{\mu\nu}$ being defined as
\[
    X_{\mu\nu}(\chi) = \Big( \nabla_{\mu}\nabla_{\nu} - g_{\mu\nu}\square \Big) G(\chi)^{-1}   - \frac{1}{2} \bigg( R\mathcal{K}(\chi) \frac{\delta\chi}{\delta g^{\mu\nu}} +
 \]
\begin{equation}
 \partial_{\kappa}\Big (R\mathcal{K}(\chi)\frac{\partial\chi}{\partial (\partial_{\kappa}g^{\mu\nu})}\Big) + \partial_{\kappa}\partial_{\lambda}\Big (R\mathcal{K}(\chi)\frac{\partial\chi}{\partial (\partial_{\lambda}\partial_{\kappa}g^{\mu\nu})}\Big ) \bigg ) \ ,
\end{equation}
with $\mathcal{K}(\chi)\equiv  \frac{2\partial{G(\chi)}/\partial{\chi}}{G(\chi)^2}$ \cite{our4}.

The functional renormalization group methods as well as other assumptions \cite{Reuter-1st, Souma}, lead to the anti-screening running gravitational coupling given by
\begin{equation}
  G(k) = \frac{G_0}{1+\omega k^2 G_0},
\end{equation}
where $k$ is the Euclidean 4-momentum, $\omega=(4/\pi)(1-\pi^2/144)$ and $G_0$ is the Newton gravity constant \cite{Bonanno & Reuter}. Hence, one can introduce the RG improvement, in the form $\omega k^2\to\xi f(\chi)$, where $\chi$ is a function of curvature invariants with dimension of length square. The function $f(\chi)\equiv \xi/\chi$ is called anti-screening running coupling since it goes to zero at the scale of very high energies ($\chi\rightarrow 0$). This behavior mimics one of the quarks and gluons, which are subject to the asymptotic freedom described by quantum chromodynamics. The scaling constant $\xi$ can be written as $G_0 \omega \xi_0 $ and $\xi_0\omega$ is a constant of the order of unity \cite{Bonanno & Reuter}.

The quantum correction term $X_{\mu\nu}$ depends on the scaling factor, and up to the first order one gets \cite{our3, our4}
\begin{equation}
  X_{\mu\nu} \simeq \nabla_{\mu}\nabla_{\nu} G(\chi)^{-1} - g_{\mu\nu}\square G(\chi)^{-1} \ ,
\end{equation}
with $\chi$ depending on a well--defined function of all independent curvature invariants such as $R, R_{\alpha\beta}R^{\alpha\beta}, R_{\alpha\beta\kappa\lambda}R^{\alpha\beta\kappa\lambda},\cdots$ \cite{our3}. However, one of the theory drawbacks is that there is no unique way to fix the form of $\chi$ \cite{Babic,Domazet-1,Domazet-2,Pawlowski}, although for non--vacuum solutions one can restrict the possible choices \cite{Babic,Domazet-1,Domazet-2,our4}.

In this direction,  the authors of Ref. \cite{Moti} study the effects of the above modifications due to Asymptotically Safe Gravity in wormholes. For this, they consider some simplifications, such as a linear Equation of State (EoS). They also study only the region very close to the throat. With this, they find that for some range of the parameters a traversable Morris wormhole is possible. However, the supposition of a linear EoS excludes a lot of possible models. Also, the study of the model with an analytical solution valid for all $r$ is important to analyze the behavior of the system through all space. The study considering these features is lacking in the literature.

In this work, we will investigate a system that provides us with both the above possibilities: a non-trivial EoS with an analytical solution for all regions of spacetime. For this, we consider the quantum effects on the Ellis-Bronnikov wormhole at the Planck scale, employing the renormalization group improved theory. Thus, in Section II, we will study models using anti-screening functions based on Ricci scalar, squared Ricci tensor, and Kretschmann scalar. We will verify that, besides the flare-out and anti-screening conditions, the principal radial energy conditions are satisfied at the Planck scale, which does not happen in General Relativity, in certain intervals of the throat radius. In addition, we will show that the quantum effects associated with asymptotically safe gravity are such that the zero tidal Ellis-Bronnikov spacetime has to be supported by a highly exotic matter, as phantom or quintessence-like energies, in a range of the wormhole throat radii and at certain regions away from the throat. In Section III, we conclude and close the paper.
\section{The Ellis-Bronnikov wormhole solution in ASG}
We consider the spherically Morris-Thorne wormhole metric given by
\begin{equation}
   d{s}^2 = e^{2\Phi(r)} d{t}^2 - \frac{d{r}^2}{1-b(r)/r} -r^2d{\Omega}_{2} \ . \label{SSM}
\end{equation}
For an anisotropic matter  $ T^{\mu}_{\nu} = \text{Diag} [\rho(r),-p_r(r),-p_l(r),-p_l(r)] \ $, the improved field equations \eqref{IEM} lead to \cite{Moti}
\begin{eqnarray}
    \kappa \rho  = &(1+f) \frac{b^{'}}{r^2} -  (1-\frac{b}{r}) (f''+\frac{2}{r} f')+ \frac{b^{'}r-b}{2r^2}f' \ ,\\
    \kappa p_r  = & -(1+f) \left(\frac{b}{r^3} -\frac{2\Phi^{'}}{r}(1-\frac{b}{r}) \right) + (1-\frac{b}{r}) \left( \Phi^{'}+\frac{2}{r}\right) f'  \ ,  \\
   \kappa p_l  = & -(1+f) \left( \frac{b'r-b}{2r^2} (\Phi'+\frac{1}{r} ) - (1-\frac{b}{r})(\Phi''+\Phi'^2+\frac{\Phi'}{r}) \right) \nonumber\\
    & + (1-\frac{b}{r}) \left(( \Phi^{'}+\frac{1}{r})f'+f''\right)  - \frac{b^{'}r-b}{2r^2}f' \ ,
\end{eqnarray}
with $\kappa=8\pi G_0$.

 For an asymptotically flat Morris wormhole, with shape and redshift functions given by
\begin{eqnarray}
b(r)=r_0^{2}/r,\\
\Phi(r)=0.
\end{eqnarray}
characterizing a zero tidal wormhole, with $r_0$ being the throat radius. The flare-out condition $b'(r_0)<1$ is readily satisfied. With this our equations simplify to
\begin{eqnarray}
\kappa\rho	&=&-(1+f)\frac{r_{0}^{2}}{r^{4}}-\frac{2}{r}(1-\frac{1}{2}\frac{r_{0}^{2}}{r^{2}})f'-(1-\frac{r_{0}^{2}}{r^{2}})f'', \label{IEQ-tt-s}
\\
\kappa p_{r}&=&-(1+f)\frac{r_{0}^{2}}{r^{4}}+\frac{2}{r}(1-\frac{r_{0}^{2}}{r^{2}})f'\ , \label{IEQ-rr-s}
\\
\kappa p_{l}&=&(1+f)\frac{r_{0}^{2}}{r^{4}}f'+(1-\frac{r_{0}^{2}}{r^{2}})\left(\frac{1}{r}f'+f''\right)+\frac{r_{0}^{2}}{r^{3}}f'\ , \label{IEQ-ll-s}
\end{eqnarray}
Now we must choose the function $f=\xi/\chi$. For the above case we have that
$$
R=-2\frac{r_{0}^{2}}{r^{4}},R^{\mu\nu}R_{\mu\nu}=4\frac{r_{0}^{4}}{r^{8}}, R_{\mu\nu\kappa\lambda}R^{\mu\nu\kappa\lambda}=12r_0^4/r^{8}
$$
therefore, in order that the condition $f>0$ is obeyed and to get the correct dimensions we must choose
\begin{equation}\label{fKretsch}
f_1=-\xi R=2\xi\frac{r_{0}^{2}}{r^{4}},f_2=\xi(R^{\mu\nu}R_{\mu\nu})^{1/2}=2\xi\frac{r_{0}^{2}}{r^{4}},f_3=\xi(R_{\mu\nu\kappa\lambda}R^{\mu\nu\kappa\lambda})^{1/2}=\xi\sqrt{12}r_0^2/r^{4}.
\end{equation}
With the above expressions we see that for the Ellis-Bronnikov wormhole we have that $f_1=f_2=f_3/\sqrt{3}$. Therefore for all cases we must be very similar behaviors. We will first consider the cases $f_1,f_2$ which are identical, and obtain analytical conclusions. At the end we plot figures for the case $f_3$.

Now we will analyze the radial energy conditions of the Ellis-Bronnikov wormhole. Such conditions are verified from the substitution of Eqs. (\ref{fKretsch}) into Eqs. (\ref{IEQ-tt-s}) and (\ref{IEQ-rr-s}). We get
\begin{eqnarray}
\kappa\rho_r&=&-\frac{r_{0}^{2}}{r^{8}}\left(-30r_{0}^{2}\xi+24\xi r^{2}+r^{4}\right),\label{rhoellis}\\
\kappa p_r&=&-\frac{r_{0}^{2}}{r^{8}}\left(-14r_{0}^{2}\xi+16\xi r^{2}+r^{4}\right),\label{prellis}\\
\kappa(\rho+p_{r})&=&-\frac{2r_{0}^{2}}{r^{8}}\left(-22r_{0}^{2}\xi+20\xi r^{2}+r^{4}\right).\label{plellis}
\end{eqnarray}
The first thing we can note is that the state equation is not linear for any value of $\xi$. Therefore, as we said in the introduction, the model considered in Ref.\cite{Moti} do not take in account the Ellis-Bronnikov wormhole, which is considered the most simple wormhole. In fact, the above $b(r)$ is never a solution to their equations. Anyway we will see that the Ellis-Bronnikov wormhole satisfies the radial energy conditions in the ASG scenario. For this we need to analyze the quantities $\rho$, $p_r$, and $\rho+p_r$, in order to verify if the null ($\rho+p_r\geq 0$), weak ($\rho\geq 0$, $\rho+p_r\geq 0$) and dominant ($\rho\geq 0$, $\rho\geq|p_r|$) energy conditions are satisfied nearby the throat.

What will determine the sign of the Eqs. (\ref{rhoellis})-(\ref{plellis}) are the terms between parenthesis. These terms are all bi quadratic equations with only two symmetric roots, respectively given by
\begin{eqnarray}
r&=&\pm\sqrt{\sqrt{6}\sqrt{\xi\left(5r_{0}^{2}+24\xi\right)}-12\xi},\label{rootrho}\\
r&=&\pm\sqrt{\sqrt{2}\sqrt{\xi\left(7r_{0}^{2}+32\xi\right)}-8\xi},\label{rootpr}\\
r&=&\pm\sqrt{\sqrt{2} \sqrt{\xi  \left(11 r_0^2+50 \xi \right)}-10 \xi }.\label{rootpl}
\end{eqnarray}
This say to us that the three quantities are positive in $r=0$ and are monotonically decreasing, changing sign in the above roots. Therefore, close to $r=0$ we can say that they are all positive, satisfying all the energy condition. However we need of the conditions close to $r=r_0$. In $r=r_0$ the above equations reduce to
\begin{eqnarray}
\kappa\rho&=&-\frac{1}{r_{0}^{4}}\left(-6\xi+r_{0}^{2}\right),\label{rhor0}\\
\kappa p_r&=&-\frac{1}{r_{0}^{4}}\left(2\xi+r_{0}^{2}\right),\label{prr0}\\
\kappa(\rho+p_{r})&=&-\frac{2}{r^{4}}\left(-2\xi+r^{2}\right).\label{plr0}
\end{eqnarray}

Therefore we arrive at some general conclusions. We see that $p_r$ must be always negative, since $p_r$ is always decreasing and at $r=r_0$ it is negative. For $\rho>0$ we see that we must add the relation $r_0^2<6\xi$. For  $\rho+p_r>0$ we must have $r_0^2<2\xi$.  Therefore we can obtain that the energy conditions depend on the relations between $r_0$ and $\xi$ as below
$$
\begin{cases}\label{energyconditionsr0}
\mbox{Null:}\quad\quad\quad\rho+p_{r}>0 & \mbox{if}\quad r_{0}^{2}<2\xi,\\
\mbox{Weak:}\quad\rho\geq 0,\rho+p_{r}\geq 0 & \mbox{if}\quad r_{0}^{2}<2\xi,\\
\mbox{Dominant:}\rho\geq 0,\rho\geq |p_{r}| & \mbox{if}\quad r_{0}^{2}<2\xi.
\end{cases}
$$
We see therefore that for $r_0^2<2\xi$ the Null, Weak and Dominant Energy Conditions are satisfied nearby the wormhole throat at the Planck scale. Therefore, a radical diference with General Relativity is that now we have the possibility that the energy conditions (\ref{energyconditionsr0}) are satisfied over the throat if $r_0^2<2\xi$. This also shows that the result of General Relativity can be recovered in the limit $\xi\to0$, as expected. From another viewpoint, in the limit $\xi\to0$, none of the conditions can be satisfied, as expected. From now on we will study the consequences of imposing that $r_{0}^{2}<2\xi$, and therefore that the energy conditions are satisfied over the throat.

First, we will investigate the presence of cosmological exotic matter in the Ellis-Bronnikov wormhole in the ASG scenario by evaluating the state parameter $\omega(r)=p_r/\rho$.  Let us first see what kind of matter is allowed at our wormhole throat. For this we have that
$$
\omega=\frac{r_0^2+2\xi}{r_0^2-6\xi}
$$
and we can analyze this as a function of $r_0$. We get easily
\begin{equation}
\begin{cases}\label{closer_0}
\mbox{Quintessence:}-1<\omega<-1/3 & \mbox{if}\quad r_{0}<\sqrt{2\xi},\\
\mbox{Phantom:}\quad\quad\quad\quad\omega<-1 & \mbox{if}\quad \sqrt{2\xi}<r_0<\sqrt{6\xi},\\
\mbox{Other Exotic Matter:}\quad\quad\quad\quad\omega>1 & \mbox{if}\quad r_0>\sqrt{6\xi}.
\end{cases}
\end{equation}

An interesting point about the above result is that General Relativity demands that $\omega=1$, but this is forbidden here since $\omega=1$ only if $\xi=0$. This again shows that the result of General Relativity can be recovered in the limit $\xi\to0$, as expected.  Now we can see the consequences of imposing $r_{0}^{2}<2\xi$, and therefore  that over the throat the energy conditions are satisfied. From Eq. (\ref{closer_0}) we see that this implies a source with $-1<\omega<-1/3$. Therefore, at $r=r_0$ our improved wormhole can satisfy the energy conditions, but must be sourced by quintessential fluid. In this case($r_{0}^{2}<2\xi$), we can also study, region by region, what are the sources that surround our wormhole. The regions where the EoS is phantom-like is given
$$
r^{4}+20\xi r^{2}-22r_{0}^{2}\xi>0.
$$
Again, this is a bi quadratic equation with two symmetric solutions given by
$$
r=\pm\sqrt{\sqrt{2} \sqrt{\xi  \left(11 r_0^2+50 \xi \right)}-10 \xi },
$$
and therefore we must have
$$
r>\sqrt{\sqrt{2} \sqrt{\xi  \left(11 r_0^2+50 \xi \right)}-10 \xi }
$$
Since we are considering $r_0<\sqrt{2}\xi$, the above expression implies that $r>\sqrt{2\xi}$. Therefore this reinforces our result that at $r=r_0$ we can never have a phantom. Now let us determine precisely the region which is Phantom-like. For this, we remember that $\rho$ has a real root where it changes sign, and $p_r$ is always negative. With this we conclude that, beyond the root (\ref{rootrho}),  we must have exotic matter with ($\omega>1$), for instance, Casimir energy ($\omega=3$).  We also conclude that the region between $\sqrt{2\xi}$ and our singularity is phantom-like. Therefore, in asymptotically safe gravity, the wormhole requires very exotic matter as source for $\sqrt{2\xi} <r<\sqrt{2\xi}\sqrt{\sqrt{51}-3}$, while in General Relativity it needs exotic matter with $\omega=1$ (stiff matter \cite{Zeldo}) at any $r$. Below we give the solution for all regions
\begin{equation}
\begin{cases}\label{regionsr}
\mbox{Quintessence:}-1<\omega<-1/3 & \mbox{if}\quad r_{0}<r<\sqrt{2\xi},\\
\mbox{Phantom:}\quad\quad\quad\quad\omega<-1 & \mbox{if}\quad \sqrt{2\xi}<r<\sqrt{2\xi}\sqrt{\sqrt{51}-3},\\
\mbox{Other Exotic Matter:}\quad\quad\quad\quad\omega>1 & \mbox{if}\quad r>\sqrt{2\xi}\sqrt{\sqrt{51}-3}.
\end{cases}
\end{equation}

Finally we consider the Kretschmann scalar. As said above the behaviour must be the same as for the other cases. Due to the similarity we only give the plot of the expressions.  In Fig. 2, we plot the quantities $\rho$, $p_r$, and $\rho+p_r$, in order to In order to visualize that the null ($\rho+p_r\geq 0$), weak ($\rho\geq 0$, $\rho+p_r\geq 0$) and dominant ($\rho\geq 0$, $\rho\geq|p_r|$) energy conditions are satisfied nearby the throat.

\begin{figure}[htb]
\centering
\includegraphics[width=0.6\textwidth]{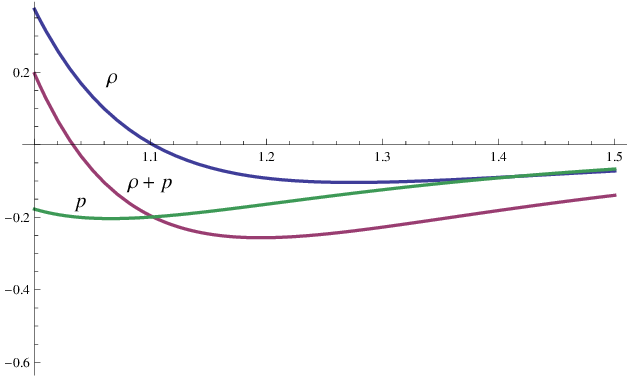}
\caption{Plot of $\rho$, $p_r$, and $\rho+p_r$, as functions of the coordinate $r$, in Planckian units, for $\xi=1$ and $r_0=1$.}
\end{figure}

Looking at Fig.1, we can notice that both Null, Weak and Dominant radial Energy Conditions are satisfied nearby the wormhole throat at the Planck scale.

 In Fig. 2 we depict the same quantities, now in the context of General Relativity ($\xi=0$).

\begin{figure}[htb]
\centering
\includegraphics[width=0.6\textwidth]{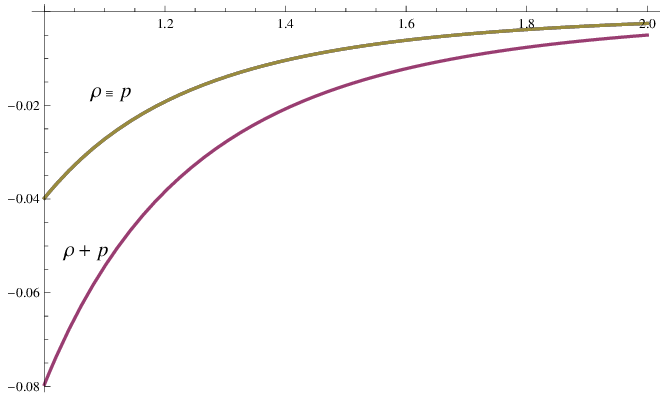}
\caption{Plot of $\rho\equiv p_r$, $\rho+p_r$, as functions of the coordinate $r$, in Planckian units, for $\xi=0$ and $r_0=1$.}
\end{figure}

Hence we conclude that, in General Relativity, the radial Null and Weak energy conditions are not satisfied nearby the throat, and the Dominant one is satisfied, and even so in the inferior limit, $\rho=|p|$.

In Fig 3 we depict the state parameter as a function of $r_0$, considering quantum improved gravity.

\begin{figure}[htb]
\centering
\includegraphics[width=0.6\textwidth]{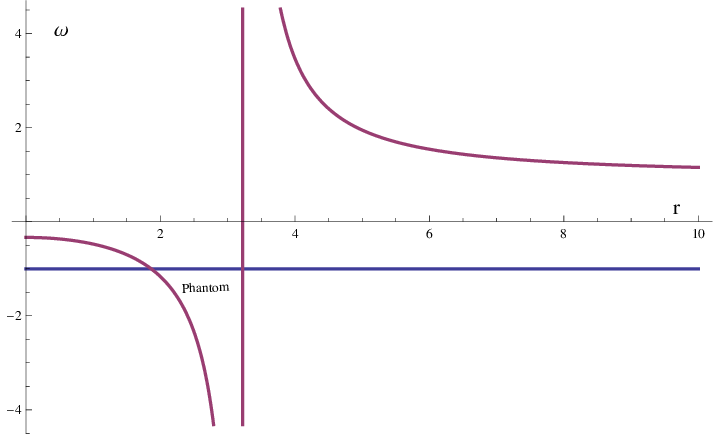}
\caption{Plot of state parameter, $\omega=p_r/\rho$, as a function of the throat radius $r_0$, in Planckian units, for $\xi=1$ and $r=r_0$.}
\end{figure}
The above figures has the same behaviors as the previous analytical expressions when we considered the functions $f_1,f_2$. Therefore the conclusions are basically the same. We should point the results for the Kretschmann scalar can be obtained just by doing $\xi\to\sqrt{4}\xi$.

\section{Conclusions}

In this paper, we have studied the presence of the simplest wormhole solution (the Ellis-Bronnikov one) in the context of Asymptotic Safety in quantum gravity, at the Planck scale. Thus, we have considered three models, which employ Ricci scalar, squared Ricci tensor, and Kretschmann scalar to perform a renormalization group improvement of the Ellis-Bronnikov wormhole of General Relativity at that scale.

In this scenario, we have checked the radial energy conditions for that wormhole solution and found that, besides both flare-out and anti-screening conditions, the Null, Weak and Dominant Energy Conditions are also satisfied nearby its throat if $r_0<\sqrt{2}\xi$. On the other hand, the wormhole obeys just the Dominant Condition in General Relativity, and only at the inferior limit, $\rho=|p_r|$.

First, we consider the Ricci scalar and squared Ricci tensor since they provide identical improvements. We analyze the improved Ellis-Bronnikov wormhole in the region very close to the throat by using $r=r_0$. We found that the effective EoS in this region forbids $\omega=1$. Therefore, the quantum gravity effects cannot match with a perfect fluid as it is considered in Ref. \cite{Moti} to generate the wormhole under consideration. In fact we consider $\omega$ as a function of $r_0$. With this, we show in Eq. (\ref{closer_0}) that the quantum gravity effects on the region nearby the wormhole throat are such that there must be a phantom, quintessence, or exotic matter with $\omega>1$ (including Casimir energy) generating thus the modified wormhole spacetime, despite the radial energy conditions are satisfied or not. The phantom-like matter, for example, is obtained for the throat radii $\sqrt{2\xi}<r_0<\sqrt{6\xi}$. However, when we impose that the wormhole does not violate the radial energy conditions over the throat, we find that $r_0<\sqrt{2\xi}$ and the only possibility in the region is given by $-1<\omega<-1/3$. Thus the quantum gravity effects require the presence of the quintessence-like matter near the throat of the wormhole.

Next, we consider the other regions of the wormhole and what kind of sources are necessary for that quantum modified gravity. Now we impose that the wormhole satisfies the radial energy conditions from the beginning. With this, we show in Eq. (\ref{regionsr}) that the wormhole can be divided into three regions when $r$ grows from the throat. The first, around the throat, must be sourced by quintessence matter-like. The second must be sourced by a phantom-like fluid and the third by exotic matter with $\omega>1$, including Casimir energy ($\omega=3$).

Third, and since the Kretschmann scalar is proportional to the Ricci scalar, we consider this case just graphically. In Fig 1 we can notice that the Null, Weak, and Dominant Energy Conditions are satisfied nearby the wormhole throat at the Planck scale. In Fig. 2 we depict the same quantities, now in the context of General Relativity ($\xi=0$). In Fig. 3 we depict the state parameter as a function of $r_0$, considering quantum improved gravity. With these plots, we can also visualize all the results of the previous cases. For example, we can see in Fig. 3 that the kind of fluid that must be present at the throat due to the quantum gravity effects can solely be quintessence-like matter, $-1<\omega<-1/3$.

Finally, we can conclude that the hypothesis of a perfect fluid is not possible in order to get traversable Ellis-Bronnikov wormholes in the context of ASG. However, at least for this case, the very exotic phantom-like matter must be present in some regions of the modified spacetime, and even at the throat for some of them, notwithstanding the non-violation of the radial energy conditions in the analyzed scenario. To study other examples with analytical solutions beyond the throat would be very important in order to verify if the phantom-like matter is a necessity for wormholes in the asymptotically safe gravity or even in more general quantum gravity scenarios. These will be the topics of our next studies.

\section*{Acknowledgements}

The authors would like to thank Conselho Nacional de Desenvolvimento Cient\'{i}fico e Tecnol\'{o}gico (CNPq) and Funda\c{c}\~{a}o Cearense de Apoio ao Desenvolvimento Cient\'{\i}fico e Tecnol\'{o}gico (FUNCAP), under grant PRONEM PNE-0112-00085.01.00/16, for the partial financial support. H.S.V. is funded by the Alexander von Humboldt Stiftung/Foundation (ID No. 1209836). This study was financed in part by the Coordena\c c\~{a}o de Aperfei\c coamento de Pessoal de N\'{i}vel Superior - Brasil (CAPES) - Finance Code 001.

\bibliographystyle{unsrt}


\end{document}